\documentclass[twocolumn]{article} 
\usepackage{hyperref}
\usepackage{graphicx}
\usepackage{xspace}
\usepackage{todonotes}
\usepackage{enumerate}
\usepackage[shortlabels]{enumitem}

\usepackage{pifont}

\newcommand{\code}[1]{\textsf{\footnotesize#1}\xspace}

\title{
The Used, the Bloated, and the Vulnerable: Reducing the Attack Surface of an Industrial Application
}

\author{%
\textsc{Serena Elisa Ponta, Wolfram Fischer, Henrik Plate, Antonino Sabetta}\\[1ex]
\normalsize SAP Security Research \\
\normalsize {\textit{\{serena.ponta,wolfram.fischer,henrik.plate,antonino.sabetta\}}@sap.com} 
}

\begin{document}

\makeatletter
\begin{figure*}
\begin{minipage}{\textwidth}
\begin{center}
{\LARGE \bf \@title}\\[4mm]
{\Large [PRE-PRINT]} \\[8mm]
\@author
\end{center}

\thispagestyle{empty}
\vspace{10mm}
\input{abstract}
\vspace{15mm}
\hrule
\vspace{10mm}
\begin{center}
{\Large Citing this paper}
\end{center}


If you wish to cite this work, please refer to it as follows:

{\tt\footnotesize
@INPROCEEDINGS\{ponta2021icsme,\\
\hspace*{3mm} author=\{Serena Elisa Ponta and Wolfram Fischer and Henrik Plate and Antonino Sabetta\},\\
\hspace*{3mm} booktitle=\{2021 {IEEE} International Conference
on Software Maintenance and Evolution (ICSME)\},\\
\hspace*{3mm} title=\{\@title\},\\
\hspace*{3mm} year={2021},\\
\}
}
\vspace{5mm}
\hrule

\end{minipage}
\end{figure*}
\makeatother

\maketitle
\thispagestyle{empty}
\pagestyle{empty}

\begin{abstract}
Software reuse may result in \emph{software bloat} when significant portions of
application dependencies are effectively unused.
Several tools exist to remove unused (byte)code from an application or its
dependencies, thus producing smaller artifacts and, potentially, reducing the
overall attack surface.

In this paper we evaluate the ability of three debloating tools to distinguish
which dependency classes are necessary for an application to function correctly
from those that could be safely removed. To do so, we conduct a case study on a
real-world commercial Java application.

Our study shows that the tools we used were able to correctly identify a
considerable amount of redundant code, which could be removed without altering
the results of the existing application tests. One of the redundant classes
turned out to be (formerly) vulnerable, confirming that this technique has the
potential to be applied for \emph{hardening} purposes. However, by manually
reviewing the results of our experiments, we observed that none of the tools can
handle a widely used default mechanism for dynamic class loading.

\end{abstract}

\section{Introduction}

The past two decades saw a considerable increase in \emph{software reuse},
particularly of \emph{open-source components}, both in commercial and free
software. Automated package management tools
allow developers to find and integrate third-party components in their projects
with minimal effort. While automated dependency management simplifies software
reuse, it may contribute to the phenomenon of \emph{software
bloat}~\cite{Soto-Valero2021}. As Gkortzis et al. put it 
``\emph{code reuse cuts both ways}'', since ``\emph{a
system can become more secure by relying on mature dependencies, or more
insecure by exposing a larger attack surface via exploitable dependencies}''~\cite{gkortzis2021reuse}.

In practice, only a fraction of the functionality (and code) of a
dependency may actually be needed, and entire components could be
redundant.
Even if some dependency code is not reachable when included in a given
application (and thus it can be considered \emph{dead code} in that context), it
can still contribute to extending the attack surface of that application, e.g.,
because it includes gadget classes leading to deserialization
vulnerabilities\footnote{\url{https://owasp.org/www-community/vulnerabilities/Deserialization_of_untrusted_data}}.

A promising way to reduce the attack surface of an application is to remove the
unused parts of its dependencies, and a number of recent publications explore
this direction proposing new techniques and tools, typically demonstrated by
applying them to open-source projects (see~Sec.\ref{sec:relwork}).
Given the potential impact of these tools in increasing the security of
enterprise applications, we conducted a case study to evaluate whether they could be adopted in practice at SAP. 

In this paper, we study a real-world commercial Java application that is part of
an SAP product, and we investigate the ability of three existing \emph{software
debloating tools} to distinguish the \emph{dependency classes} that are used
from those that 
could be removed without compromising the correct behaviour of the application.
We propose a methodology to evaluate \emph{(i)} how the removal of the classes
reported as redundant impacts attack surface of the bundled application and
\emph{(ii)} how this affects the correct execution of the application. 

Each tool was able to report a considerable number of classes as redundant. Once
removed, the existing application tests continue to pass. We detected a
(formerly) vulnerable class among those removed, which is an example of a small
but tangible reduction in the attack surface.
A manual review of the classes identified as redundant, however, revealed that
none of the tools we considered was able to identify a class that is dynamically loaded at runtime,
and that has been confirmed by the developer as being required.

The remainder of the paper is structured as follows.
Sec.~\ref{sec:selected_tools} provides an overview and comparison of (the
selected) state-of-the-art debloating tools. Sec.~\ref{sec:study} introduces the
case-study methodology, and summarizes its results. Sec.~\ref{sec:relwork}
summarizes related work, and Sec.~\ref{sec:conclusion} concludes the paper
briefly outlining possible future work.


\section{Debloating Tools}
\label{sec:selected_tools}

\begin{table*}[ht]
    \centering
 \caption{Summary of tool characteristics}
\begin{center}
\begin{tabular}{ |c||c|c|c|c| }
\hline
Characteristic & DepClean & Maven Shade & ProGuard \\
 \hline\hline
 \shortstack{Original \\use-case} & \shortstack{Remove dependency \\declaration (Maven projects)} & \shortstack{Create self-contained \\ Uber-Jar (Maven projects)} & \shortstack{Shrink and obfuscate Java \\archives (Maven-independent)} \\
 \hline
 Approach & \shortstack{Bytecode analysis\\(starting from prj. classes,\\considers literals \\ to cover reflection)} & \shortstack{Bytecode analysis\\(starting from \\prj. classes)} & \shortstack{Bytecode reachability analysis\\(starting from entry points)} \\
 \hline
 \shortstack{Slicing \\Granularity} & \shortstack{Java archives\\(class-level info available \\ as debug info)} & Java classes & \shortstack{Java class members} \\
 \hline
 \shortstack{Analysis \\input} & Compiled project classes & Compiled project classes & \shortstack{Compiled project classes and \\ classes (members) specified \\ as entry-points} \\
 \hline
 \shortstack{Analysis \\output} & \shortstack{Modified POM file\\(with removed/excluded \\ dependencies)} & \shortstack{Uber-Jar with \\needed classes \\(unmodified bytecode)} & \shortstack{Uber-Jar with needed classes\\(bytecode potentially \\shrinked and obfuscated)} \\
 \hline
\end{tabular}
\label{tab:tool-table}
\end{center}
\end{table*}

In our study, we consider mature, widely-used open source tools, as well as open source tools
for software debloating readily available, sufficiently documented, and easily
applicable to Java applications that use Apache Maven as build
system\footnote{\url{https://maven.apache.org/}}.
Table~\ref{tab:tool-table} summarizes and compares the main characteristics of
the three open-source tools.

\noindent\textbf{Apache Maven
Shade}\footnote{\url{https://maven.apache.org/plugins/maven-shade-plugin/}}
(Maven Shade for short) is a mature and well-established plug-in for Maven that
creates self-contained Java archives (\emph{Uber-Jars}), to be used at
application runtime. Uber-Jars include all the application classes as well as
all classes of the (runtime and compile time) dependencies. 

As of version~1.4, Maven Shade supports the minimization of Uber-Jars such that
only classes actually required for the artifact are re-bundled (option
\texttt{minimizeJar}). The set of needed classes is computed using
jdependency\footnote{\url{https://github.com/tcurdt/jdependency}}, which uses
ASM\footnote{\url{https://asm.ow2.io/}} to search the bytecode for referenced
classes.

\noindent\textbf{ProGuard}\footnote{\url{https://github.com/Guardsquare/proguard}} is a
widely adopted obfuscator and shrinker for Java and Kotlin (Android) applications. It
is typically applied to mobile applications to reduce download times and protect
intellectual property via obfuscation.

Given Java archives and the specification of entry points as input, ProGuard
recursively identifies the classes and class members that can be reached. Many
other configuration options enable and fine-tune additional ProGuard features,
(such as, field or method removal, obfuscation, method inlining, class merging).

\noindent\textbf{DepClean}~\cite{Soto-Valero2021} identifies and removes bloated
dependencies that are part of the dependency tree of the project under
analysis, but whose code is not used (neither directly nor indirectly) by the application. Differently from Maven
Shade and ProGuard, the focus of DepClean is not to produce a compact Java archive
for use at runtime, but rather \emph{to simplify the dependency tree at development
time}. Moreover, it is meant to work at the granularity of entire Java archives.
The tool can be configured to produce detailed information about used code at the
granularity of Java classes (option \texttt{createResultJson}), 
which makes it possible to evaluate its ability to identify the classes
required by the application.

DepClean extends the Apache Maven Dependency Analyzer and uses ASM for bytecode
analysis in order to build a Dependency Usage Tree, which extends the standard
Maven dependency tree with edge labels to indicate whether direct, transitive
and inherited project dependencies are used or bloated respectively. DepClean
also parses the constant pool table of Java class files to cover dynamic,
reflection-based invocations done through string literals and string
concatenations.


\section{Case-study}
\label{sec:study}

Our case study investigates the ability of existing debloating tools to minimize
the dependencies of an industrial Java application without breaking it. In
particular we consider the ability of the tools presented in
Sec.~\ref{sec:selected_tools} to identify the code required by the
application at class level. As ProGuard also supports 
shrinking at finer granularity to remove class members, we consider this tool in
two flavours: ProGuard$^m$, shrinking used classes at member level, and
ProGuard$^c$, leaving used classes untouched. Finally, we focus on the effect
that the size reduction has on the attack surface of the application. 

\subsection{Methodology}
\label{sec:meth}
To compare the existing debloating tools, we use the following methodology to 
apply them to Maven projects.

\noindent\textbf{Vanilla execution.} We build and test the application, without
using any debloating tool, in order to collect information required for the
comparison.
The dependencies of Maven projects are specified in a \code{pom.xml} file and
have a \emph{scope} that determines the phase of the build process in which they
are required.
As we focus on the reduction of code required in production,
we consider the scopes \code{compile} and \code{runtime} as target of the
debloating tools. Consistently with previous literature~\cite{bruce2020jshrink},
we execute existing tests and we use their results as a proxy for \emph{semantic
preservation}.

Concretely, the vanilla execution:
\begin{itemize}
    \item[(1)] Ensures that the application successfully builds with all tests passing (\code{mvn install} succeeds);
    \item[(2)] Collects all test cases, all application class names, and all
    \code{compile} and \code{runtime} dependencies with the class names therein;
    \item[(3)] Detects the presence of vulnerable classes.
\end{itemize}

\noindent\textbf{Tool execution.} Our investigation targets the ability of the
tools to identify all and only the dependency code required by the application.
As a result, we perform the debloating step outside of the tools, and rely on
them only for providing the set of required classes.
Accordingly, the tool execution comprises the following steps:
\begin{itemize} 
	\item[(1)] Run DepClean, Maven Shade, ProGuard$^m$, ProGuard$^c$. 
	\item[(2)] Transform the tool output into a file containing the list of used classes (if not already available).
    \item[(3)] Collect the names of all classes of \code{compile} and \code{runtime} dependencies reported by the tools as used by the application.
	\item[(4)] Copy those classes from the output of the tool to \texttt{target/classes}.
	\item[(5)] Adjust the \texttt{pom.xml} to remove all \code{compile} and \code{runtime} dependencies.
	\item[(6)] Run the existing tests on the debloated application.
	\item[(7)] Detect the presence of vulnerable classes in the debloated application.
\end{itemize}

The list of used classes (cf. step (2)) from Maven Shade, ProGuard$^m$, and
ProGuard$^c$ was created by listing the content of the Jar artifact produced by
the tool. For DepClean, we enabled the configuration setting
\code{createResultJson} and rewrote the class names contained in the result file
into a plain list in order to have the same output for each tool. Also note that
for ProGuard$^m$ and ProGuard$^c$ we had to create an application-specific
configuration file containing the information of all application and test
classes to be used as entry points for the analysis (\code{keep} option) and we
disabled all optimization and obfuscation features. 

As the shrinking option of ProGuard always removes unused methods of a class
unless a configuration forces the tool to keep it untouched, in the case of
ProGuard$^m$ we run the tool as-is. For ProGuard$^c$, to get the list of used
untouched classes, we iterate steps (1) and (2) by adding the used classes as
additional entry points (to be kept as-is) in the configuration file until no
additional class is reported as used.

In step (4) we copy the classes reported as used to the project's
\texttt{target/classes} folder so that they will be available when running the
existing test, i.e., they are treated as application classes, and in step (5) we
remove \code{compile} and \code{runtime} dependencies from the \texttt{pom.xml}.
We opted for this custom debloating as it allows us to focus on the ability of
the tools to identify the used classes while allowing us to uniformly collect
the details required for measuring the reduction in terms of size and attack
surface.

The vulnerable open-source classes were obtained analysing the fix commits for
open-source vulnerabilities available at \url{https://github.com/SAP/project-kb}
(the up-to-date dataset from~\cite{pontaMSR}).

\subsection{Subject Application}

For our case study, selected a Maven project that is part of SAP's Energy Data Management solution. It
uses JAXB and EclipseLink\footnote{\url{https://www.eclipse.org/eclipselink/}}
to (un)marshal XML documents related to energy measurements. It is actively
developed and several releases have already been made available to customers
through the deployment on the SAP Cloud Platform. The project is characterized as
follows:
\begin{itemize}
\item 10 direct dependencies (2 compile, 2 provided, 6 test)
\item 20 resolved dependencies (4 compile, 3 provided, 13 test)
\item 260 application classes, 2725 compile dependency classes
\item 62 test classes amounting to 446 test cases
\end{itemize}
 
\subsection{Results}
The methodology of Sec.~\ref{sec:meth} was applied to the application above.
We had a successful run of the vanilla execution and of the tools execution on
Ubuntu 18.04 using JDK 1.8.0 , Maven 3.8.1, Maven Shade 3.2.4,
proguard-maven-plugin\footnote{\url{https://wvengen.github.io/proguard-maven-plugin/}}
2.3.1 (configured to use ProGuard version 7.0.1), and DepClean created from
revision \texttt{cbfc395} in \url{https://github.com/castor-software/depclean}.

\begin{table*}
    \centering
 \caption{Results of vanilla and tool executions}
    \label{tab:results}
\begin{center}
\begin{tabular}{ |c||c|c|c|c| }
\hline
Execution & Classes & Size (KB) & Test success & Vulnerable classes  \\
 \hline\hline
 Vanilla & 2725 & 15033 & 446 & 1\\
 DepClean & 11 & 57,26 & 446 & -\\
 Maven Shade & 12 & 57,63 &  446 & -\\
 ProGuard$^m$ & 1 & 4 &  446 & -\\
 ProGuard$^c$ & 11 & 57,26 &  446 & -\\
 
 \hline	
\end{tabular}
\end{center}
\end{table*}

Table~\ref{tab:results} shows the results of the executions. In the vanilla execution we collected 2725
classes from the 4 compile dependencies of the application, amounting to 14,68
MB of disk space (cf. Column "Size"). Both Depclean and ProGuard$^c$ reported
the same set of 11 classes as being used. Maven Shade reported one additional
class. ProGuard$^m$ was able to reduce the dependencies to a single class by
removing all members not used by the application, which contained all the
references to the 10 classes reported by ProGuard$^c$, DepClean and Maven Shade.
As a large share of classes were reported as redundant, the size on disk was
significantly reduced in all cases. For all the tools, the existing tests were
still passing on the debloated application we constructed as described in
Sec.~\ref{sec:meth}.

\begin{table*}
    \centering
 \caption{Reduction in the number of classes for each compile dependency}
    \label{tab:deps-size}
\begin{center}
\begin{tabular}{ |c|c||c|c|c|c|c| }
\hline
Dependency (Maven artifactId) & Scope & Classes & DepClean & Maven Shade & ProGuard$^c$ & ProGuard$^m$ \\
 \hline\hline
 commons-io & direct & 182 & 11 & 11 & 11 & 1\\
 org.eclipse.persistence.moxy & direct & 264 & - & 1 & - & -\\
 org.eclipse.persistence.core & transitive & 2075 & - & - & - & -\\
 org.eclipse.persistence.asm & transitive & 204 & - & - & - & -\\
 \hline
 Total & & 2725 & 11 & 12 & 11 & 1\\

 \hline
\end{tabular}
\end{center}
\end{table*}

Table~\ref{tab:deps-size} details, for each compile dependency, how many classes
the tools report as used by the application. With the constraint of leaving the
original classes unmodified, all tools identify as used the same set of 11
classes from \code{commons-io}. Instead, ProGuard$^m$ shrinks unused members
from a used class of \code{commons-io} thus removing the 10 classes imported
therein. 
Maven Shade is the only tool reporting a \code{package-info.class} from
\code{org.eclipse.persistence.moxy} (\code{moxy} for short) as used. It
contains a single runtime annotation applicable to other classes of the same
package. However, no other class is reported as used, thus, it would not be
applied to any class after debloating.

By manual inspection, we observed that the application makes use of a service
implementation offered by the direct dependency \code{moxy}, declared
according to the Java SPI (Service Provider Interface) mechanism. Java SPI
allows service consumers to only reference the service interface, while the
actual implementation is made available at runtime. None of the tools was able
to identify the class specified in the Java SPI configuration file, thus no
service implementation would be available.

Finally, we observed that one (formerly) vulnerable class was part of the application
dependencies, and has been removed by all debloating tools (cf. column
"Vulnerable Classes" of Table~\ref{tab:results}). The class is
\code{org.apache.commons.io.FilenameUtils.java}, contained in \code{commons-io},
and subject to CVE-2021-29425\footnote{Fixed by
\url{https://github.com/apache/commons-io/commit/2736b6f}.}.

\subsection{Discussion}
Despite having a considerable number of test cases,
the case-study shows the limitations of tests as an oracle for semantics
preservation.
The application developer confirmed that the unavailability of the
service implementation class, reported as redundant by all tools, would
break the application at runtime.

ProGuard allows manual configuration of entry points to be kept, so we run the
tool specifying the SPI service implementation class as additional entry point.
As a result, 209 classes of the direct dependency \code{moxy} are reported as
used, as well as 34 and 1340 of its transitive dependencies \code{asm} and
\code{core} (full Maven artifactId available in Table~\ref{tab:deps-size}).
Thus, considering also the results of our manual inspection, the application
dependencies could be reduced by half, still removing the (formerly) vulnerable
class and reducing its attack surface. 

Our case study shows the potential of debloating dependencies on an industrial
application of limited size and complexity (e.g., just four dependencies required
at runtime) and already points out a critical need for improvement.

\section{Related Work}
\label{sec:relwork}

The effect of software reuse on security is investigated by Gkortzis et al.
in~\cite{gkortzis2021reuse}, who show empirical evidence of the relation between
the size of a code base and its likelihood to contain some vulnerabilities.
Recently, Soto-Valero et al. conducted a large-scale study to assess the
prevalence of bloated dependencies in the Maven
ecosystem~\cite{Soto-Valero2021}. In the same paper, they presented DepClean,
one of the tools we used in our case study.
In \cite{bruce2020jshrink}, Bruce at al. propose JShrink, a framework to debloat
Java application using static and dynamic analysis techniques. Though their
implementation is only available as replication package, the results they
reported show the potential of the approach.

A related problem is how to measure the actual security improvements obtained by
debloating: one approach is to count the number of known vulnerabilities
(CVEs) removed~\cite{qian2020slimium}.
Other works, for example~\cite{bruce2020jshrink}, use metrics based on the
number of \emph{gadget chains} that can be successfully removed.

This paper investigates well-known and readily available debloating tools to
evaluate their ability to discriminate used from redundant code at class level.
This is done at the case of an industrial application. Moreover, it quantifies
the attack surface reduction in terms of removed vulnerable classes.


\section{Conclusion}
\label{sec:conclusion}

Considering the successful debloat of \texttt{commons-io}, including the removal
of a (formerly) vulnerable class, the case-study confirms the potential of
debloating tools to reduce an application's attack surface. It also shows that
state-of-the-art tools could not handle a widely-used, standard Java mechanism
for dynamic class loading as used in the application at hand.

Future work should consider additional debloating tools and techniques to
evaluate how they deal with dynamic features. Besides, to use debloating tools
in industrial settings, they must handle software of increasing size and
complexity, and integrate easily in CI/CD pipelines, with limited manual
configuration.

\smallskip\small\noindent\textbf{Acknowledgements.}
{\skip0=\baselineskip\small\baselineskip=\skip0 This work is partly funded
by EU grants No.~952647 ({\sc AssureMOSS}) and No.~830892 (\textsc{Sparta}).\normalsize}


\end{document}